\documentstyle[11pt,newpasp,twoside,epsf]{article}
\def\edcomment#1{\iffalse\marginpar{\raggedright\sl#1\/}\else\relax\fi}
\reversemarginpar

\def\nheh{\hbox{n$_{\rm He}$/n$_{\rm H}$}}

\def\Mdot{\hbox{$\dot {M}$}}

\def\Lsun{\hbox{\it L$_\odot$}}

\def\Msun{\hbox{\it M$_\odot$}}
\def\Minit{\hbox{\it M$_{\rm initial}$}}
\def\Msunyr{\hbox{\it M$_\odot\,$yr$^{-1}$}}

\def\Vinf{\hbox{$v_\infty$}}
\def\kms{\hbox{km$\,$s$^{-1}$}}

\def\simgr{\mathrel{\hbox{\rlap{\hbox{\lower4pt\hbox{$\sim$}}}\hbox{$>$}}}}
\def\simls{\mathrel{\hbox{\rlap{\hbox{\lower4pt\hbox{$\sim$}}}\hbox{$<$}}}}
\def\HeI{He\,{\sc i}}
\def\arcsec{\hbox{$^{\prime\prime}$}}

\begin{document}
\title{Stellar Collisions and Mergers in the Galactic Center}
 \author{Donald F. Figer}
\affil{Space Telescope Science Institute, 3700 San Martin Dr., Baltimore, MD 21218; 
Johns Hopkins University, 3400 N. Charles St., Baltimore, MD 21218}
\author{Sungsoo S. Kim}
\affil{UCLA, 405 N. Hilgard Ave., LA, CA 90095}

\begin{abstract}
Stars are most likely to merge or collide in regions of the highest stellar densities, and
our own Galactic
Center contains many stars packed into a relatively small volume -- even the ambient
stellar number density in the central 50~{\it pc} is quite high, $\sim10^3~stars~pc^{-3}$.
More striking, the three compact young clusters in this region have central densities 
as high as $10^6~stars~pc^{-3}$. We discuss these extreme environments and the possibility
that stellar mergers and collisions have recently occured there. In particular, we predict that at least one
massive star in the Arches cluster has already experienced a stellar merger in its short
lifetime. Further, the Pistol Star, in the nearby Quintuplet cluster, might owe
its apparent relative youth to a rejuvinating stellar merger. Finally, the apparently young
stars in the central arcsecond could be products of either collisions, inducing atmospheric stripping, or 
mergers.
\end{abstract}

\section{Introduction}
The Galactic Center (GC) is a very dense stellar environment and was recognized early on as 
a region which might contain products of stellar mergers and collisions. Soon after the
identification of the dense stellar cluster in the GC (Becklin et al.\ 1978), 
Lacy, Townes, \& Hollenbach (1982) suggested that material in the ionized gas in the GC mini-spiral 
was produced by star-star collisions, but it is now thought that the mini-spiral traces
gas pulled in from a wayward molecular cloud, or the nearby circumnuclear ring. Nonetheless,
stellar collisions and mergers may still play an important role in describing phenomena in
the central parsec. Indeed, a new favorite target for this possibility is the tight cluster of
stars in the central arcsecond (Bailey \& Davies 1998; Alexander 1999), as its high
velocity stars appear (Genzel et al.\ 1997; Ghez et al.\ 1998) to be very 
young (Eckart et al.\ 1999; Figer et al.\ 2000).

There are, at least, two other interesting regions, as regards stellar collisions, 
just 30~{\it pc}, in projection, from the center: the Arches cluster (Nagata et al.\ 1995; Cotera et al.\ 1996;
Serabyn, Shupe, \& Figer 1997; Figer et al.\ 1999a), and the Quintuplet cluster (Okuda et al.\ 1990;
Nagata et al.\ 1990; Glass, Moneti, \& Moorwood 1990; Figer, McLean, \& Morris 1999b). They are 
nearly identical to each other, and to the young cluster in the central parsec, except for age 
-- the Arches cluster is $\sim$2~Myr old, half the age of the other two clusters. 

This paper is an introduction to the GC, and objects therein, as it pertains
to studies of stellar collisions and mergers. We start by introducing the environment of the
GC and finish with 3 case studies of potential mergers or collisions.

\section{An Introduction to the Galactic Center for Stellar Merger/Collision Astronomers}
For this paper, the GC refers to the central 100~{\it pc} wherein the central mass
density approximately follows a power law, with slope $\sim-$2 (Becklin \& Negebauer 1968; 
Lindqvist et al.\ 1992). The enclosed mass
within r$\sim$~50~{\it pc} is $\sim~3(10^8)~M_{\sun}$ (Lindqvist et al.\ 1992), and the
stellar volume density is estimated to be $>5(10^5)~stars~pc^{-3}$ at r~=~1~{\it pc}. 
In addition to the molecular gas and largely old stellar population, there is ample evidence
of star formation in the GC. 

Radio data reveal many embedded HII regions which likely harbor massive stars that are currently forming.
Indeed, the Sgr A West Mini-spiral, Sickle, and Thermal Arched Filaments are tracers for gas 
ionized by spectacular young stellar clusters: 1) the central cluster, 2) the Quintuplet
cluster, and 3) the Arches cluster. The central cluster has attracted the most attention
over the past 30 years, being the first discovered (Becklin \& Negebauer 1968). 

The Central cluster contains over 30 evolved massive stars having \Minit\ $>$~20~\Msun
(Forrest et al.\ 1987; Allen et al.\ 1990;
Krabbe et al.\ 1991; Krabbe et al.\ 1995; Eckart et al.\ 1995; Libonate et al.\ 1995; Blum et al.\ 1995;
Genzel et al.\ 1996; Tamblyn et al.\ 1996). A current
estimate of the young population includes $\approx$10 WR stars, 20 stars with Ofpe/WN9-like 
{\it K}-band spectra, several red supergiants, and
many luminous mid-infrared sources in a region of 1.6~pc in diameter centered on Sgr~A$^*$
The spatial distribution of the early- and late-type stars has been the subject of several different studies
(Becklin \& Neugebauer 1968, Allen 1994, Rieke \& Rieke 1994, Eckart et al.\ 1995, Genzel et al.\ 1996).
The continuum surface distribution maps reveal that early-type stars concentrate towards the center, i.e. in
the IRS16 sources, but red giants and supergiants are mainly concentrated outside of 
a 5\arcsec\ region near the center. 

Najarro et al.\ (1994) and Najarro et al.\ (1997) have modelled a half dozen members of the non-WR
hot stellar population in the center, finding that they are generally ``Ofpe/WN9''
stars with strong winds ($\Mdot\sim$ 5 to 80 $\times 10^{-5}\, \Msunyr$) and
relatively small outflow velocities (\Vinf $\sim$ 300 to 1,000~\kms).
The effective temperatures of these objects were found to range from 17,000\,K to 30,000\,K
with corresponding  stellar luminosities of 1 to $30 \times 10^{5}$ \Lsun.
These results, together with the strongly enhanced helium abundances (\nheh\ $>$~0.5), indicate that
the \HeI\ emission line stars power the central parsec and belong to a young stellar cluster of massive stars
which formed a few million years ago.

The Quintuplet Cluster is equally impressive for its stellar content, containing over 
30 stars having \Minit\ $>$~20~\Msun, including 4 WC types, 4 WN types, 2 LBVs, and many OBI stars
(Figer, McLean, \& Morris 1999b). First noted for its five bright infrared sources (Okuda et al.\ 1990,
Nagata et al.\ 1990, Glass et al.\ 1990), the cluster is now known to contain thousands of stars
(Figer et al.\ 1999a; see Figure 1). Of them, the Pistol Star (LBV) is the most luminous, L$\sim10^{6.6}~\Lsun$, and
star \#362 (LBV) in Figer, McLean, \& Morris (1999b; Geballe et al.\ 2000) is a close second.
Given the mix of spectral subtypes for the most luminous stars, Figer, McLean, \& Morris (1999b) estimate
a cluster age of 3$-$5~Myr, although significant age differences remain, i.e.,
the Pistol Star is thought to be $\approx$ 2 Myrs old (Figer et al.\ 1998).

The Arches cluster contains a dense collection of emission-line stars (Nagata et al.\ 1995; 
Cotera et al. 1996), and several thousand fainter members within a half-light radius of $\approx$0.2 pc
(Figer et al.\ 1999a). In fact, this cluster is the densest and one of the most massive young
clusters in the galaxy, having a central density $>5(10^5)~\Msun~pc^{-3}$, and a total mass
$>~10^4~\Msun$.

In general, the three Galactic Center clusters are young ($<5$~Myr), compact ($<1$~pc), 
and appear to be as massive as the smallest
Galactic globular clusters ($\sim10^4~\Msun$). However, these compact
young clusters have several interesting dynamical characteristics
that distinguish them from globular clusters: 1) they have
short dynamical timescales ($t_{dyn} \sim
10^{5-6}$~yr and $t_{rh} \sim 10^{6-7}$~yr, respectively); 2) they are situated
in strong tidal fields (the tidal radius of $10^4 \Msun$ cluster located
30~pc from the Galactic center is 
$\sim 1$~pc); and 3) mass segregation may occur on a timescale shorter
than the lifetimes of the most massive stars in these clusters, i.e.\ massive
stars may play an important role in the dynamical evolution of the cluster.

\begin{figure}
\vskip 0.75in
\plotfiddle{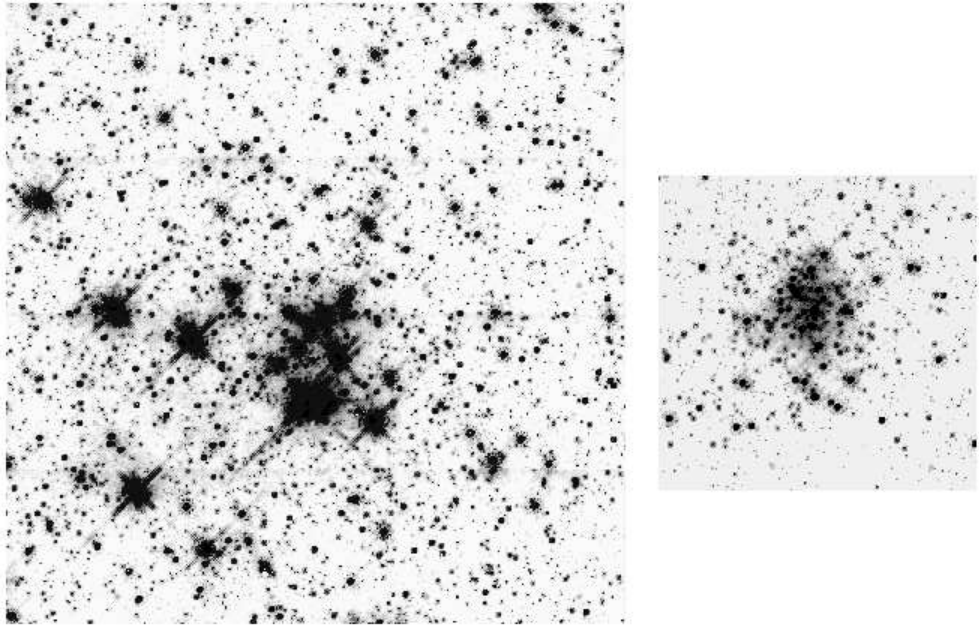}{2.5in}{-90}{50}{50}{-200pt}{260pt}
\vskip .2in
\caption{HST/NICMOS images of the Quintuplet ({\it left}) and Arches ({\it right}) custers (Figer et al.\ 1999a).}
\end{figure}

\section{Three Galactic Center Testbeds for Studying Mergers and Collisions}

\subsection{The Pistol Star - Is It the ``Smoking Gun''?}

The ``Pistol Star,'' located roughly at the center of curvature of the Pistol HII
region in the Quintuplet cluster, is one of the most luminous and massive stars
in the Galaxy (Figer et al.\ 1998; see Figure 2). Its initial mass, $\sim150~\Msun$,
is so great, that Figer et al.\ (1998) predict a very short lifetime of $\sim2~Myr$,
yet most of the other stars in the Quintuplet are twice as old. The crossing time in
the cluster is $\sim10^5~yr$, and the cluster has already made a complete orbit 
around the Galactic Center, assuming a circular orbit. In addition, its natal molecular
cloud is nowhere to be seen. Given these facts, we assume that all of the stars in
the cluster formed at roughly the same time. This begs the question as to how the
Pistol Star, with its extreme high mass, can still be burning so bright after
4~Myr. 

Figer et al.\ (1998) proposed that the star may have had an unusual formation history. 
They suggest that the star has very little angular momentum, which perhaps allowed its
mass to grow so much, consequently, having 
less-than-average mass-loss towards the cool side of the HR diagram.
During core hydrogen burning, it arrived at its Eddington limit
and strongly increased its mass-loss rate. While this scenario is possible, it
requires an age in the range 1.7$-$2.1 Myrs.

Another possible explanation relies on a stellar merger to create the Pistol Star. 
In this scenario, the Pistol Star is the result of a merger between two very massive
stars, or, at the very least, a result of accretion-induced collisions between
star-forming cores (Bonnell et al.\ 1998). 

\begin{figure}
\vskip 0.75in
\plotfiddle{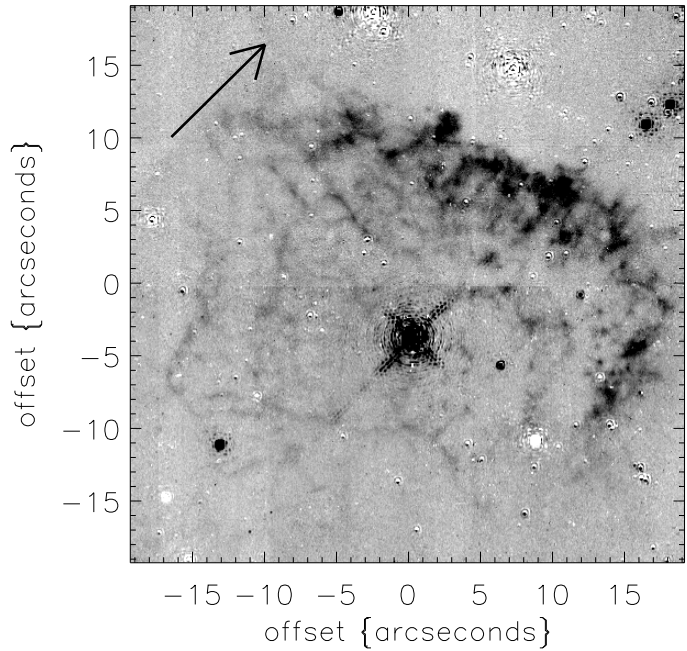}{2.5in}{0}{70}{70}{-170pt}{-260pt}
\vskip .2in
\caption{Pa-alpha image of Pistol Star and its surrounding nebula.}
\end{figure}

\subsection{Collisions and Mergers in the Quintuplet and the Arches Clusters}

The fact that we currently observe only two young clusters
near the GC (except the central cluster right at the GC) raises a
natural question about the lifetimes of these compact young clusters (CYCs).
Using anisotropic Fokker-Planck (F-P) models, Kim, Morris,
\& Lee (1999; KML hereafter) surveyed lifetimes of CYCs for various
initial mass functions (IMFs), cluster masses ($M$), and Galactocentric
radii ($R_g$), and found that clusters with $M \la 2 \times 10^4 \,
M_\odot$ and $R_g \la 100$~pc evaporate in $\la 10$~Myr.  These
unparalleled, short evaporation times ($t_{ev}$) of CYCs are first due to
short $t_{dyn}$ and $t_{rh}$, and strong tidal forces, but the mass loss
accompanying the evolution of massive stars is also responsible for shortening
$t_{ev}$ of clusters that last longer than $\sim 3$~Myr.

F-P models are statistical models involving distribution functions.
It is the statistical stability and fast computing time of the F-P
models that the survey-type study by KML required.  Takahashi
\& Portegies Zwart (1998) found a good agreement between anisotropic
F-P models and N-body simulations for globular clusters by adopting an
``apocenter criterion'' and an appropriate coefficient for the speed
of star removal behind the tidal radius (KML adopted these as well).
However, some extreme conditions of CYCs may be inconsistent with the
assumptions inherent to F-P models.  As discussed in KML, the conditions
required by F-P models, $t_{dyn}~\la~t_{rh}$ and $t_{dyn}~\la~t_{se}$
($t_{se}$ is the stellar evolution timescale) may be violated, especially
in the core at certain epochs.  Moreover, the active participation of
a large mass range of stars in the dynamics, which is another peculiarity
of CYCs, is difficult to realize in F-P models that embody a mass
spectrum with a restricted number (usually 10-20) of discrete mass
components.  A greater number of components would better express the
mass spectrum, but then the number of stars in each component would
become smaller. Since F-P models basically assume an infinite
number of stars for each component, they overestimate the
role of a component that has too few stars in it.
On the contrary, too small a number of components would not properly
realize the whole mass spectrum.  By fixing the number of stars
in the most massive components, KML tried to carefully account
for a relatively small number of the most massive stars.  

CYCs, estimated to have $\sim 10^{3-4}$ stars, are one of few systems
for which real-number N-body simulations are feasible on current
workstations.  Among many virtues of N-body simulations, the natural
realization of the mass spectrum and the tidal fields is particularly
beneficial to the study of CYCs.  These benefits make N-body
simulations treat mass segregation and evaporation of stars exactly,
thus providing better density profiles and mass spectra as a function of
radius.  These are photometric observables, and are partially available
from the HST/NICMOS observations of the Arches and Quintuplet by
Figer et al.\ (1999a; partial availability is owed to
crowding at the cluster center and to the detection limit at the faint end
of the luminosity function).  The Arches and Quintuplet are estimated
to be only $\sim 2$ and $\sim 4$~Myr old, respectively
(Figer et al.\ 1999a), but their currently observed structures
must already have deviated from the initial ones due to their
rapid dynamical evolution.  With both real-number N-body simulations and HST
observations of the two clusters in hand, one has a rare chance of deriving
not only the current characteristics of these systems, but their initial
conditions as well.

We performed a series of N-body simulations
1) to compare the lifetimes of CYCs with those from F-P models
obtained by KML, 2) to find the initial cluster conditions
that best match the current observations of CYCs,
and 3) to test the effects of initial mass segregation and different initial
density profiles on dynamical evolution of the CYCs.

Kim et al.\ (2000) investigated the dynamical evolution of the Quintuplet and
Arches clusters using Aarseth's Nbody6 codes. The simulations used a realistic
number of test points (5,000--20,000), and thus represent an
improvement over the Fokker-Planck (F-P) models (KML) which
surveyed a variety of cluster lifetimes and initial conditions. In general,
both the N-body and F-P models indicate that clusters with a mass $\la 2 \times 10^4 \, M_{\sun}$
evaporate in $\sim 10$~Myr. In addition, the N-body simulations were used
with the observed projected number density profiles
and stellar mass functions from {\it HST}/NICMOS
observations by Figer et al.\ (1999a) to infer the initial conditions of the 
Arches cluster: a tidal radius of 1~{\it pc}, total mass of $2 \times 10^4 \, M_{\sun}$, and mass function 
slope of $-$0.7 (versus $-$1.35 for Salpeter). We also find that the lower stellar mass limit,
the amount of initial mass segregation, and the choice of initial density
profile (King or Plummer models) do not significantly affect the dynamical
evolution of CYCs. 

The number of events in the cluster core may be calculated by 
\begin{equation}
	\int N_c n_c \sigma_{event} v_c dt,
\end{equation}
where $N_{\rm c}$, $n_{\rm c}$, and $v_{\rm c}$ are number of stars, number density, and velocity
in the core, and $\sigma_{\rm event}$ is the cross section of a certain event.
$N_{\rm c}$ is roughly constant over time and does not vary much from cluster to cluster
(from CYCs to globular clusters).  We know that
$n_{\rm c}$ and $v_{\rm c}$ of CYCs are similar to that of globular clusters (except for
the Quintuplet which is thought to be in its final disruption phase).
    The largest difference comes from $\sigma_{event}$ which depends mostly
    on the size (in case of tidal interaction) of stars at the high density
    region.  While the massive stars in globular clusters disappear before
    the corecollapse takes place, those in CYCs survive until the epoch of
    central density peak due to very short relaxation time.  Since a
    significant mass segregation takes place during the first 1 Myr of
    the CYCs due to an exceptionally wide mass spectrum, the mean mass
    at the core quickly becomes very large.  Thus $\sigma_{event}$ at the center
    of CYCs will be signicantly larger than that of globulars.
 But, the number of merger
events in our most recent n-body calculations for the Arches (IMF slope~$=\Gamma=-0.7$,
initial mass~$=2(10^4)$~\Msun) is only a few (1-5). This is basically because
the cluster does not live long (with high central densities). In any case, several runs 
predict a merger which results in a total mass larger than 200~\Msun.
This may exlain the existence of the Pistol Star.

If we put the Arches cluster to a place where
the tidal radius would be 10 times larger than the current value,
i.e.\ farther away from the Galactic center, the lifetime of the
cluster would be 100 times longer.  But this does not mean we would
observe 100 times more merger events because the cluster would
live with relatively small central densities for most of its lifetime.
This is due to the rapid expansion of the core after 2~{\it Myr} when
the stellar evolution stars.  So, the number of events would be only
slightly larger than at the current location of the Arches.

Note that our comparison between the N-body simulation and
HST observations suggests a relatively flat IMF ($\Gamma=-0.7$).  Massive stars
have larger radii and thus larger collision cross sections.  So those who
do hydrodynamical simulations of close stellar encounters might
consider encounters between very massive stars. We refer the 
reader to Kim \& Lee (1999) for theoretical cross sections of tidal capture
events between two stars with a large mass contrast.

\subsection{The Central Parsec and the Stellar Cluster Near the Black Hole}

The Central cluster has drawn the most attention regarding stellar collisions and
mergers for obvious reasons, given its high stellar density and the region's strong gravitational
field which produces high stellar velocities. From an observational point of view, Sellgren et al.\ (1990) 
noted an absence of bright red giants near the central
5\arcsec, suggesting that stellar collisions might have disrupted the outer
layers of such stars. While it is true that bright red giants are preferentially located
outside the central few arcseconds, Figure 4 in Genzel et al.\ (1996) and Figure 2 in 
Figer et al.\ (2000) clearly show that spectra
of background flux in the central few arcseconds exhibits CO absorption bandheads suggestive
of K giants. The observational evidence, then, for collision-induced disruption of stellar
atmospheres is strongest for the brightest red giants, for which Genzel et al.\ (1996) 
calculate that a core density of $10^{6.5}~\Msun~pc^{-3}$ would be sufficient to generate
the requisite number of star-star interactions; however, the impact parameter they used was
too large to effect a complete ejection of the red giant atmosphere. 
Alexander (1999) revisited this issue, considering close interactions which could lead to the depletion of 
red giant atmospheres in the central few arcseconds. He concluded that it is possible to
explain the lack of bright red giants in that region due to close encounters, but that this
conclusion is critically dependent on a cusp-like distribution of those stars, i.e.\ a
density of $>5(10^7)~\Msun~pc^{-3}$. Bailey \& Davies (1999) performed a similar simulation,
finding that stellar collisions are unlikely to have produced the apparent depletion of
red giants in the center 5\arcsec. In particular, they find that the collision rate is
likely to have been too low. A final
determination on this issue will await higher resolution imaging which would sample the distribution
of stars in the center. 

Another candidate for products of stellar mergers/collisions in the center lies in the dense
cluster of stars in the central arcsecond (Genzel et al.\ 1997; Ghez et al.\ 1998).
Genzel et al.\ (1997) first claimed that some of these dozen or so stars are similar in their
K-band spectral morphology and brightnesses to OB main sequence stars. Using much higher
spectral resolution, Eckart et al.\ (1999) and Figer et al.\ (2000) arrived at similar
conclusions. Such stars only remain on the main sequence for 20~Myr, so they must have
formed quite close to their current location; however, bound cores so near to the central
black hole require extremely high densities, $>10^{11}~cm^{-3}$. In addition, OB main
sequence stars have relatively high masses, $\sim~20~\Msun$, so the protostellar cores
must have been very massive. An alternate possibility relies on stellar collisions to
produce objects which only appear to be so hot. However, this is unlikely, given that the
same requirements discussed above for the red giants pertain. In addition, it is questionable
that a stripped red giant would have such large luminosities.

\section{Conclusions}

In conclusion, we find that the three young stellar clusters in the Galactic Center are
prime proving grounds for stellar merger and collision theories. In particular, these
clusters are the densest and most massive young clusters in the Galaxy. They have individual
members which might be products of stellar mergers, i.e.\ the Pistol Star, and they
may contain products of near-misses, or collision, i.e.\ in the Central cluster.


\begin{references}
\reference Alexander, T. 1999, {\it ApJ}, {\bf 527}, 835
\reference Allen, D.A., 1994, in ``The Nuclei of Normal Galaxies,'' eds.\ R. Genzel \& A. I. Harris (Dordrecht: Kluwer), 293
\reference Bailey, V. C., \& Davies, M. B. 1999, {\it MNRAS}, {\bf 308}, 257
\reference Becklin, E. E., \& Negebauer, G. 1968, {\it ApJ}, {\bf 151}, 145
\reference Blum, R. D., DePoy, D. L. \& Sellgren, K. 1995, {\it ApJ}, {\bf 441}, 603
\reference Bonnell, I. A., Bate, M. R., \& Zinnecker, H. 1998, {\it MNRAS}, {\bf 298}, 93
\reference Cotera, A. S., Erickson, E. F., Colgan, S. W. J., Simpson, J. P., Allen, D. A., \& Burton, M. G. 1996, {\it ApJ}, {\bf 461}, 750
\reference Eckart, A., Genzel, R., Hofmann, R. Sams, B. J., \& Tacconi-Garman, L. E., 1995, {\it ApJ}, {\bf 445}, L26
\reference Eckart, A., Ott, T., \& Genzel, R. 1999, {\it A\&A}, {\bf 352}, L22
\reference Figer, D. F., Becklin, E. E. , McLean, I. S. , Gilbert, A. M., Graham, J. R., Larkin, J. E., Levenson, N. A., Teplitz, H. I., Wilcox, M. K., \& Morris, M., 2000, {\it ApJ}, {\bf 533}, L49 
\reference Figer, D. F., Kim, S. S., Morris, M., Serabyn, E., Rich, R. M., \& McLean, I. S. 1999a, {\it ApJ}, {\bf 525}, 750 
\reference Figer, D. F., McLean, I. S., \& Morris, M. 1995, {\it ApJ}, {\bf 447}, L29
\reference Figer, D. F., McLean, I. S., \& Morris, M. 1999b, {\it ApJ}, {\bf 514}, 202
\reference Figer, D. F., McLean, I. S., \& Najarro, F. 1997, {\it ApJ}, {\bf 486}, 420.
\reference Figer, D. F., Najarro, F., Morris, M., McLean, I. S., Geballe, T. R., Ghez, A. M., \& Langer, N. 1998, {\it ApJ}, {\bf 506}, 384
\reference Figer, D. F., Rich, R. M., Morris, M., McLean, I. S., Serabyn, E., Puetter, R., \& Yahil, A. 1999c, {\it ApJ}, {\bf 525}, 759
\reference Forrest, W. J., Shure, M. A., Pipher, J. L., \& Woodward, C. A. 1987, in ``The Galactic Center,''
ed. D. Backer, AIP Conf. Proc. 155, 153
\reference Geballe, T. R., Najarro, F., \& Figer, D. F. 2000, {\it ApJ}, {\bf 530}, 97
\reference Genzel, R., Eckart, A., Ott, T., \& Eisenhauer, F. 1997, {\it MNRAS}, {\bf 291}, 219
\reference Genzel, R., Thatte, N., Krabbe, A., Kroker, H. \& Tacconi-Garman 1996, {\it ApJ}, {\bf 472}, 153
\reference Ghez, A. M., Klein, B. L., Morris, M., \& Becklin, E. E. 1998, {\it ApJ}, {\bf 509}, 678
\reference Glass, I. S., Moneti, A. \& Moorwood, A. F. M. 1990, {\it MNRAS}, {\bf 242}, 55P
\reference Kim, S. S., Figer, D. F., Lee, H. M., \& Morris, M. 2000, {\it ApJ}, {\bf December 20, 2000}
\reference Kim, S. S., \& Lee, H. M. 1999, {\it A\&A}, {\bf 347}, 123
\reference Kim, S. S., Morris, M., \& Lee, H. M. 1999, {\it ApJ}, {\bf 525}, 228
\reference Krabbe, A., Genzel, R., Drapatz, S. \& Rotaciuc, V. 1991, {\it ApJ}, {\bf 382}, L19
\reference Krabbe, A., et al.\ 1995, {\it ApJ}, {\bf 447}, 95
\reference Libonate, S., Pipher, J. L., Forrest, W. J., Ashby, M. L. N. 1995, {\it ApJ}, {\bf 439}, 202
\reference Okuda, H., Shibai, H., Nakagawa, T., Matsuhara, H., Kobayashi, Y., Kaifu, N., Nagata, T., Gatley, I. \& Geballe, T. R. 1990, {\it ApJ}, {\bf 351}, 89
\reference Nagata, T., Woodward, C. E., Shure, M., Pipher, J. L. \& Okuda, H. 1990, {\it ApJ}, {\bf 351}, 83
\reference Nagata, T., Woodward, C. E., Shure, M, \& Kobayashi, N. 1995, {\it AJ}, {\bf 109}, 1676
\reference Najarro, F., Hillier, D. J., Kudritzki, R. P., Krabbe, A.,
Genzel, R., Lutz, D., Drapatz, S. \& Geballe, T. R. 1994, {\it A\&A}, {\bf 285}, 573
\reference Najarro, F., Krabbe, A., Genzel, R., Lutz, D., Kudritzki, R. P., \& Hillier, D. J. 1997,
{\it A\&A}, {\bf 325}, 700
\reference Sellgren, K., McGinn, M. T., Becklin, E. E., \& Hall, D. N. B. 1990, {\it ApJ}, {\bf 359}, 112
\reference Serabyn, E., Shupe, D., \& Figer, D. 1998, {\it Nature}, {\bf 394}, 448
\reference Takahashi, K., \& Portegies Zwart, S. F. 1998, {\it ApJ}, {\bf 503}, L49
\reference Tamblyn, P., Rieke, G. H., Hanson, M. M., Close, L. M., McCarthy, D. W., Jr., \& Rieke, M. J. 1996, {\it ApJ}, {\bf 456}, 206
\end{references}
\end{document}